\documentstyle[12pt,epsf]{article}
\catcode`\@=11 \@addtoreset{equation}{section}  %%
\renewcommand{\theequation}                     %%
         {\arabic{section}.\arabic{equation}}   %%
\title{Quantum Field Theory on Fock Projective Space}
\author{P. Leifer }
\date{School of Physics and Astronomy \\ Raymond and Beverly Sackler 
Faculty of Exact Sciences \\ 
Tel-Aviv University, Tel-Aviv 69978, Israel}
\begin{document}
\maketitle
\begin{abstract}
It is shown that some analog of the ``second quantization''
exists in the framework of $CP(N)$ theory.
I analyse conditions under that ``geometrical bosons''
may be identified with real physical fields.
The compact character of a state manifold should
preserve the quantities of dynamical variables
from divergences. 
\end{abstract}
\vskip .2cm
\section{Introduction}
Quantum field theory (QFT) is now the basis of our knowlege
about microscopical dynamical processes. The essential
part of the apparatus of this theory is the ``second
quantization'' method (SQM) developed by Dirac 
\cite{Dirac1,Dirac2}. This approach is so powerful and 
prolific that can not be overestimated \cite{Weinberg,TFD}.
Notwithstanding there are limits for both general
foundations of standard QFT and SQM as well. These
difficulties were  subjected to analysis 
by numerous authores (see \cite{Barut} and citetions
therein). Three directions of Dirac's
attempts to overcome these obstacles are interesting 
for us. These are:

a) Attempts to avoid divergences
by using Heisenberg pictures \cite{Dirac2};

b) The costruction  of extended model of an electron
\cite{Dirac3};

c) Efforts to obtaine Coulomb ineraction
between charges without interaction term
\cite{Dirac4}.

Here I will discuss only introduction of ``geometrical
bosons'' in association with the direction a).

I would like to emphasize two fundamental assumptions 
which  lie in the basis of Dirac approach. They are 
the flat (and fixed) character of spacetime structure and
the method of classical analogy \cite{Dirac1,Dirac2}.
The second one leads to well known form of free
quantum Hamiltonians for electromagnetic field and electrons, 
and for the energy operator of interaction 
between them. The first assumption seems to be 
absolutely natural when we are enough far from the 
problem of quantum gravity. But as it has been
emphasized by Dirac, the quantum electrodynamics can not
be consistent theory because at short distances
there are processes of creation and annihilation of
new kinds of quantum particles \cite{Dirac2}. For such
particles we, of course, have not classical analogy
and, therefore, one has not a correct Hamiltonian
for interacting fields just in the dangerous
area of our theory. Futhermore,  
spacetime description of the deep inelastic processes
requires {\it dynamical treatment of spacetime structure} \cite{Le1}.

In order to understand the root of problem
it will be convenient to returne to the
regularization procedure in the framework of Dirac  
approach \cite{Dirac2}. Here I will preserve 
Dirac's notations. The second order equation
for operator-value state in the representation
of interaction is as follows
\begin{equation}
i\frac{\partial K^*_2}{\partial t}=
\frac{e^2}{\pi}\sum_n \psi_n\{Y_{ni}+Z_{ni}\}
e^{i(E_n-E_i)t} + Add. terms,
\label{K2}
\end{equation}
where $Y_{ni}$, for example, is expressed by the formula
\begin{equation}
Y_{ni}=(4\pi)^{-1}\int <n|\alpha^\mu e^{i({\bf kx})}
\frac{\nu+(E/|E|)}{|{\bf k}|+\nu (E_n - E_i)}
e^{-i({\bf kx})} \alpha_\mu |i> \frac{d^3k}{|{\bf k}|}.
\label{Y}
\end{equation}
This expression is subjected to some of the methods of
regularization which have been discussed by Dirac
in detailes. Let me to point out some weak
points of this approach besides that what was 
mentioned by Dirac:

A.{\it Only for enough large regularization 
parameter $g$ differences between results of
regularization do not depend on $g$}.

B.{\it One needs a local (in time) vacuum state
vector}.

Dirac has assumed that a state vector more or less closed
to ordinary Schr\"odinger state vector does not
exist at all because it does not belong to any Hilbert 
space at each instant of time. {\bf I think that there
exists some state vector as a tangent vector to
some smooth manifold which (vector) creeps along this 
curved state space and, therefore, belongs to different
tangent Hilbert spaces  at each instant of time}. 
That is a quantum interaction (like in general relativity (GR)) 
connected with a curvature, but with the  curvature
of a state space, not spacetime. In order to take
into account all possible kinds of interactions
one should use the general properties of quantum
interaction. In the framework of the unitary
quantum dynamics and the concept of the deformation 
of quantum states the most natural structure of the 
interaction is the coset structure
$SU(N+1)/S[U(1)\times U(N)]$ \cite{Le1}.
Therefore the simplest model of such generalized quantum 
interaction is represented by the internal dynamics in
$CP(N)$ \cite{Le1,Le2,Le3,Le4,Le5}.
The question is: does this model contain interaction
which may be identified with real physical
interactions and what is the positive the role
of the compactness of $CP(N)$?

\section{Absolute Space of Conserved Quantities}
Dirac used in (\ref{K2}) $\psi_n$ as 
\begin{equation}
\psi_n=\int <n|{\bf x}a> \psi_{ax}d^3x,
\label{psin}
\end{equation}
where $<n|{\bf x}a>$ is eigenfunction of the equation
\begin{equation}
[ \alpha_r(-i \frac{\partial}{\partial x_r}
+eA_r)+\alpha_m m-eA_0]_{ab}<{\bf x}b|n> 
=E_n <{\bf x}a|n>.
\label{En}
\end{equation}
Here Dirac assumed that $\psi_n$ are so-called q-numbers
which obeys anticommutation relations since he
delt with fermions. 
This step leads to singular functions and we should try to
avoid such ``forcible'' quantization and to work with
invariant (geometrical) properties of the state space.

Hereafter we will assume that our $\Psi^n$ are 
ordinary c-numbers describing entanglement of all 
possible degrees of freedom and all real physical 
states correspond to the 
variations of relative local coordinates (\ref{Pi}). 
The realization of a nonlinear version of quantum
field theory requires an essentially new costruction
of the both state space and spacetime \cite{Le1,Le4,Le5}.
Let us begin with the construction of the state space.
During this reconstruction we should preserve all
robust propertias of the standard QFT models
which we have in the relativistic QFT of free fields
\cite{Weinberg,TFD}. It is well known just the 
switching of interactions leads to major difficulties.
I think that geometrical model of interaction which
has been mentioned above demands unification 
relativity and quantum theory in a totally non-standard
manner. 

In this model we will assume that there is
{\it absolute space of conserved quantities} (ASCQ) 
which is (from the mathematical point of view) the 
Fock's Hilbert space which is constructed
with help of $[0]$-multitude by the action on the 
vacuum  vector $|V>$ \cite{TFD}. But one can not 
interpret this Fock space in the ordinary sense
as the space of state of ``particles''. All quantum
particles are dynamical systems. But ASCQ is only ``reservoir 
of conserved quantities'' (or ``charges'') and it has not a 
dynamical content. 
It serves merely for the enumeration of ``charges'' and, 
therefore, helps us to choose initial conditions and 
identification of symmetries of ``elementary particles''. 
For instance quarks are not particles as some isolated entities.  
In this space there is a {\it vacuum state} which 
does not contain any conserved quantities. This is
the state $|V>=|0,...,0,...>$ with zero occupation 
numbers for any conserved ``charge''. In that sense
the vacuum is ``the universal breakdown of conservation lawes, 
i.e. interaction (or maybe ``chaos'') in a pure form''. 
We will assume  that the modulus of  the vacuum state is 
defined by the formula $<V|V>=R_{vac}^2$ and the radius 
$R_{vac}$ connected with some fundamental interaction constant 
(e.g. fine structure constant 
$\alpha=\frac{e^2}{\hbar c}$).
Then creation from the vacuum state some conserving
``charge'' (or some self-consistend set of ``charges'',
 i.e. particle) leads to the existence of conservation lawes.
There are two kinds of perturbation of the vacuum 
state:

a) Unitary perturbations which generated coherent
states by transformation from the coset manifold
$G/H=SU(N+1)/S[U(1)\times U(N)]$:

b) Non-unitary multifold creation of identical quantum 
particles in different points of spacetime.

Pursuing to understand b) from the point of view of a) 
we will assume that  
it is possible in the tangent fiber bundle
over projective Hilbert space $CP(N)$.
The structure of each tangent space is isomorphic
to Fock's Hilbert space.  This isomprphism is in fact 
a simple, but it needs a clarification. 

In some sense
ASCQ is like absolute Newtonian space and time
where ``in'' and ''out'' free fields have
a physical sense. In this space ``external'' field
has physical meaning as well; frequences of 
oscillations in these ``external'' fields, wave
vectors, masses of ``elementary particles'' etc.,
are presented. 
In ASCQ there is the universal ``world time''
of  St\"ueckelberg-Horwitz-Piron (see \cite{Horwitz}
and citation therein) in the 
sense of the {\bf omnipresence of the coinsidence 
of the periods  of physically identical 
quantum processes}. The problem is to calculate
this  ``world time'' from the first principles.
But it is impossible in the framework of free
fields model and one should go toward interacting
fields. 

I would like to emphasize that  {\bf as the kinematical 
comparison of periods in moving frames in special relativity 
destroys the ``independence'' of periods (absolute
Newtonian time) as a dynamical ``comparison''  
(an interaction) destroys independence of dynamical
variables in different quantum setup}. Our aim 
is to find the geometric origin of this
mutual influance. 

Let us introduce the separable Fock space 
\begin{equation}
{\cal{H}}=\{|\Psi>=\sum _a^{\infty} \Psi^a |a>;
\sum_a^{\infty} |\Psi^a|^2 < \infty \},
\end{equation}
(here our $|a>$ is identical with $\xi_a$ 
\cite{TFD}). We will use a density submanifold
\begin{equation}
D=\{ \sum _{a=0}^N \Psi^a |a>; N=N_1+...+N_p < \infty \},
\end{equation}
where $p$ is the numbers of kinds of the ``charges''.
In order to avoid difficulties with the Dirac problem
of divergences a), which mathematically are reflected
by the existance of the unitary nonequivalent 
representations of commutation relations
(see \cite{TFD} and citations therein), we will use
the compact projective space $CP(N)$ by the 
compactification of Fock's Hilbert space ${\cal{H}}$. 
We will assume that there is 
the ``sphere of vacuum states'' generated by 
transformations from the coset 
$G/H=SU(N+1)/S[U(1)\times U(N)]$
\begin{equation}
||V||^2=\sum_{a=0}^N|V^a|^2=R_{vac}^2
\label{R}
\end{equation} 
for any unitary equivalent vacuum vector
\begin{equation}
|V>=\sum_{a=0}^{N}V^a|a>,
\label{V}
\end{equation} 
where $\{|a>\}_0^N$ corresponds to some complete set
on the finite dimensional space we start with,
and the $V^a$ are normalized to a scale parameter
$R_{vac}$, which controls the curvature 
of the K\"ahler's projective metric space of state.
Then for any unitary equivalent state vector from $\cal{H}\rm $
\begin{equation}
|\Psi>=\sum_{a=0}^{N}\Psi^a|a>,
\label{Psi}
\end{equation} 
for which
\begin{equation}
||\Psi||^2=\sum_{a=0}^N|\Psi^a|^2=R_{vac}^2 
\label{mod}
\end{equation} 
one can define coordinates, which we shall call
{\it local projective coordinates}
for example in the chart $U_0$ ($\Psi^0 \neq 0$)
\begin{equation}
\Pi^1=R_{vac}\frac{\Psi^1}{\Psi^0},\quad \Pi^2=R_{vac}
\frac{\Psi^2}{\Psi^0},...,
\Pi^i=R_{vac}\frac{\Psi^i}{\Psi^0},...,
\Pi^N=R_{vac}\frac{\Psi^N}{\Psi^0},
\label{Pi}
\end{equation}
and since, then, 
\begin{equation}
R_{vac}^2=||\Psi||^2=|\Psi^0|^2(1+
R_{vac}^{-2}\sum_{i=1}^N|\Pi^i|^2),
\label{mod1}
\end{equation} 
we can choose the phase of $\Psi^0$ shuch that
\begin{equation}
\Psi^0=\lambda R_{vac},\quad 
\Psi^1=\lambda \Pi^1,
\Psi^2=\lambda \Pi^2,...,
\Psi^N=\lambda \Pi^N,
\label{pspi}
\end{equation}
where ($1 \leq i \leq N$) and
\begin{equation}
\lambda(R_{vac},\Pi)=\frac{R_{vac}}{\sqrt{\sum_{s=1}^N |\Pi^s|^2+R_{vac}^2}}.
\label{lambda}
\end{equation}

Now we can express homogeneous coordinates $\Psi$ in local
coordinates $\Pi$:
\begin{equation}
\Psi^0=\frac{R_{vac}^2}{\sqrt{\sum_{s=1}^N |\Pi^s|^2+R_{vac}^2}},...,
\quad 
\Psi^i=\Pi^i \frac{R_{vac}}{\sqrt{\sum_{s=1}^N |\Pi^s|^2+R_{vac}^2}}.
\label{PsiPi}
\end{equation}
It is easy to evaluate ($a=0$)
\begin{equation}
\frac{\partial \Psi^0}{\partial \Pi^i}=-\frac{1}{2}
\frac{R_{vac}^2 \Pi^{*i}}{\left(\sqrt{\sum_{s=1}^N |\Pi^s|^2
+R_{vac}^2}\right)^3},
\frac{\partial \Psi^0}{\partial \Pi^{*k}}=-\frac{1}{2}
\frac{R_{vac}^2 \Pi^{k}}{\left(\sqrt{\sum_{s=1}^N |\Pi^s|^2
+R_{vac}^2}\right)^3}
\end{equation}
and for other components ($a \geq 1$) one has
\begin{eqnarray}
\frac{\partial \Psi^m}{\partial \Pi^i}     & = &
R_{vac}\left(\frac{\delta^m_i}{\sqrt{\sum_{s=1}^N |\Pi^s|^2
+R_{vac}^2}}-
\frac{1}{2} \frac{\Pi^m \Pi^{*i}}
{\left(\sqrt{\sum_{s=1}^N |\Pi^s|^2+R_{vac}^2}\right)^3}\right), 
\nonumber \\
\frac{\partial \Psi^{*m}}{\partial \Pi^{*k}} & = &
R_{vac}\left(\frac{\delta^m_k}{\sqrt{\sum_{s=1}^N |\Pi^s|^2
+R_{vac}^2}}-
\frac{1}{2} \frac{\Pi^{*m} \Pi^{k}}
{\left(\sqrt{\sum_{s=1}^N |\Pi^s|^2+R_{vac}^2}\right)^3}\right). 
\end{eqnarray}
Therefore one can express infinitesimal invariant interval
in the original Fock Hilbert space as followes
\begin{equation}
\delta L^2= \delta_{ab}\delta \Psi^a \delta \Psi^{*b}
= G_{ik*}\delta \Pi^i \delta \Pi^{*k}=\sum_a \frac{\partial 
\Psi^a}{\partial \Pi^i}
\frac{\partial \Psi^{*a}}{\partial \Pi^{*k}} 
\delta \Pi^i \delta \Pi^{*k}.
\end{equation}
That is the metric tensor of the original flat
Hilbert space in the local coordinates $\Pi$ is 
\begin{eqnarray}
G_{ik*}=\sum_{a=0}^N \frac{\partial \Psi^a}{\partial \Pi^i}
\frac{\partial \Psi^{*a}}{\partial \Pi^{*k}} =
R_{vac}^2 \frac{(\sum_{s=1}^N |\Pi^s|^2+R_{vac}^2)\delta_{ik}-\frac{3}{4}
\Pi^{*i}\Pi^k}{(\sum_{s=1}^N |\Pi^s|^2+R_{vac}^2)^2}.
\label{FSmet}
\end{eqnarray}

One can evaluate the Poisson brackets of the components $\Psi^a,\Psi^b$
\begin{equation}
\{\Psi^a,\Psi^b\}=\frac{\partial \Psi^a}{\partial \Pi^i}
\frac{\partial \Psi^{*b}}{\partial \Pi^{*i}}-\frac{\partial 
\Psi^{*a}}{\partial \Pi^{*i}}\frac{\partial \Psi^{b}}{\partial \Pi^{i}}
\end{equation}
in two particular cases; the first one is
\begin{equation}
\{\Psi^0,\Psi^k\}=
\frac{R_{vac}^3}{2(\sum_{s=1}^N |\Pi^s|^2+R_{vac}^2)^2}
(\Pi^{k}-\Pi^{*k}),
\end{equation}
and the second one is as follows
\begin{equation}
\{\Psi^i,\Psi^k\}= 
\frac{R_{vac}^2 \sum_{s=1}^N |\Pi^s|^2}{4(\sum_{s=1}^N 
|\Pi^s|^2+R_{vac}^2)^3}
(\Pi^{i}\Pi^{*k}-\Pi^{*i}\Pi^k).
\end{equation}
It is easy to see that all our expressions are 
smooth fuctions of the local projective
coordinates. At first sight they have neither
connection with spacetime (or momentum) representation
which we need for a dynamical description.
This representation one should obtain by the
introduction of a {\it dynamical spacetime}
\cite{Le4}.

\section{Geometrical Bosons}
In our description the real deviation from the 
vacuum state related to rate of the changing of local
vacuum (tangent vector fields). These deviation
(deformation) one can identify with the state vector
of some quantum system. In that sense our formalism
corresponds on the one hand to Dirac formalism 
\cite{Dirac2} because tangent vectors are differential 
operators \cite{KN},
but on the other hand they are vectors of some tangent
Hilbert space, and correspond to ordinary 
Schr\"odinger formalism.

Now one can introduce ``geometrical bosons'' assuming
that 
\begin{equation}
[\frac{\partial}{\partial \Pi^k},\Pi^i]_-=\delta^i_k.
\label{comm}
\end{equation}
In order to agree the standard Fock representation
and the definition of vacuum state by a holomorphic 
function $F_{vac}$ we should introduce the ``Hamiltonian''
of {\it geometrical bosons} as the tangent vector fields
\begin{equation}
\Xi^i_k(bos)=\hbar \omega \Pi^{*i}\frac{\partial}{\partial \Pi^{*k}}.
\label{bos}
\end{equation}
Since $\frac{\partial F_{vac}}{\partial \Pi^{*k}}=0$
one has:
\begin{eqnarray}
\Xi^i_k(bos)F_{vac}=0; \cr
\Xi^i_k(bos)\Pi^{*s} F_{vac}=\hbar \omega 
\Pi^{*i}\delta^s_k F_{vac}; \cr
\Xi^i_k(bos)\Pi^{*s_1} \Pi^{*s_2}F_{vac}=\hbar \omega 
\Pi^{*i}(\Pi^{*s_2}\delta^{s_1}_k +
\Pi^{*s_1}\delta^{s_2}_k)F_{vac}; \cr
. \cr
. \cr
. 
\label{excit}
\end{eqnarray}
One can introduce the function of excitations
of different degrees of freedom 
$F(s_1,...,s_{\cal{N}})=\Pi^{*s_1} \Pi^{*s_2}...
\Pi^{*s_{\cal{N}}}F_{vac}$ and the function of a 
multifold excited degree of freedom
$F(s;\cal{N})$ $=(\Pi^{*s})^{\cal{N}}F_{vac}$.
It is easy to see that only in very particular 
case we have the coinsidence with the equidistant
spectrum of harmonic oscillator. These are
\begin{eqnarray}
\Xi^i_k(bos)F(s;1)=\hbar \omega F(s;1); \cr
\Xi^i_k(bos)F(s;2)=2\hbar \omega F(s;2); \cr
. \cr
. \cr
. \cr
\Xi^i_k(bos)F(s;{\cal{N}})
={\cal{N}} \hbar \omega F(s;{\cal{N}}).
\label{oscill}
\end{eqnarray}
Therefore one can think
that besides ordinary bosons (e.g. photons)
the model of the ``geometrical bosons'' containes
the different kinds of excitations with a
non-equidistant spectrum. This is the 
consequence of the multiplete (anisotropic)
structure of an interaction.
  
In order to identify such kinds of excitations
with real physical fields we should know whether the
character of spacetime propagation or dispersion
law of these geometrical bosons. Self-consistent
solution of this problem requires the introduction
of {\it dynamical spacetime on the basis of a quantum
setup} \cite{Le4}. {\it Self-consistent here means that
spacetime coordinates are function of the relative
quantum amplitudes}. In accordance with this ideology 
the quantum ``internal'' degrees of freedom have 
a primacy in a comparison with ``external'' spacetime
coordinates of a position. For simple models of 
quantum mechanics like oscillator or single electron
in the field of a nucleus we have matrices corresponding
to operators of spatial coordinates
\begin{equation}
{\bf x}_{ik}=\int \psi^*_i{\bf x}\psi_k d^3x
\label{x}
\end{equation}
and we can solve eigenproblem in order to find state
vector $|{\bf x}_n>$. Then it is possible to
introduce a wave function 
$\Psi({\bf x})=<{\bf x}_n|\Psi^n>$ which gives
us the spatial destribution of system.
These results are based on
the method of classical analogy where we have
some the spacetime model of a quantum system
(we know what is ${\bf x}$).
But now we have to use only geometrical--field 
model of dynamical
process in a quantum state space and spacetime
coordinates should be expressed as functions of
pure quantum dynamical variables, dipole moment
of transition, for example \cite{Le4}.  Then one 
suspects that our argument $\Pi^{*s}$ of $F(s;{\cal{N}})$
may be identified with the relative amplitude
which has been obtained as a result of the ansatz of 
``squeezing'' described in \cite{Le4}.

The energy variation which associated with 
infinitesimal gauge transformation of the local 
frame  with the coefficients of the connection
is
\begin{eqnarray}
\delta H= \frac{1}{\mu}\delta U=\frac{1}{\mu} A_m \delta \Pi^m 
=\frac{1}{\mu}\frac{\delta U}{\delta \Pi^m}\delta \Pi^m
=-\frac{\hbar}{\mu} \Gamma^i_{km}\xi^k \frac{\partial \Psi^a}
{\partial \Pi^i}\delta \Pi^m |a>.
\label{Am} 
\end{eqnarray}
It may be described by the form of the curvature
of $CP(N)$
\begin{eqnarray}
\omega^i_k= \Gamma^i_{km}\delta \Pi^m, 
\label{omega} 
\end{eqnarray}
for which one has the transformation law of tangent
vector fields
(boson fields in paricular case, as well) in a
tangent Hilbert space
\begin{eqnarray}
\xi^{'i}=U^i_m \xi^m
\label{transXi} 
\end{eqnarray}
and then
\begin{eqnarray}
\omega^{'i}_k=  U^{-1i}_m \omega^{m}_n U^n_k
+\delta U^{-1i}_t U^t_k.
\label{transOm} 
\end{eqnarray}
This local (in $CP(N)$) non-Abelian gauge field 
(until now it is a latent 
field because we have not a spacetime dependence) at 
the level of Fourier components which takes the place
of some universal physical field carriering
interactions. Dynamical description of this gauge
field requires the ``internal'' intoduction of 
spacetime coordinates in pure quantum manner.
It looks like Wilczek--Zee
gauge potential \cite{WZ} but it has, of course, 
different physical meaning. The physical status of 
this interaction is subjected to our study. 

One of the important case of linear transformations
in tangent space is tranformation defined by the 
tensor of the curvature of Jacobi 
$K_{\vec{\xi}}(\vec{J})=R(\vec{\xi},\vec{J})\vec{\xi}:
T_\Pi CP(N) \to T_\Pi CP(N)$ \cite{Milnor}.
This is defined by the sectional curvature in 2-direction
between tangent vector $\vec{\xi}$ to a geodesic and
the transversal Jacobi vector field $\vec{J}$. 
It has already been emphasized \cite{Le5} that gauge 
transformations connected with variation (rotation) of 
geodesics as whole. They are generated by transformations
from the isotropy group of the vacuum state
$H=U(1) \times U(N)$. This geodesic variation (as well
as any geodesic variation) may be described by the Jacobi vector field.
The equation for this vector field has in the case 
$CP(N)$ very simple form since $CP(N)$ is 
homogeneous (and even symmetric) space
\begin{equation}
\frac{d^2 J_i}{ds^2}+\kappa_i J_i=0,
\label{Jacobi}
\end{equation}
and very simple oscillation solutions
\begin{equation}
J_i=c_i \sin \sqrt{\kappa_i} s
\label{sol}.
\end{equation}
We will use affine arclength parametrization
$ds=\frac{E}{\hbar} dt$. Then one can rewrite 
(\ref{Jacobi}) if 
$\frac{\partial E}{\partial t}=\frac{\partial E}{\partial s}
=0$ as follows
\begin{equation}
\frac{d^2 J_i}{ds^2}=\frac{\hbar^2}{E^2}\frac{d^2 J_i}{dt^2}= -\kappa_i J_i.
\label{Jac}
\end{equation}
That is in the natural parametrization the sectional
curvature is defind by the formula
\begin{equation}
\kappa_i=\frac{\hbar^2 \omega_i^2}{E^2}
\end{equation}
and one see that we have automatic (geometrical)
quantization of the gauge field connected with
the geodesic variation. It means that {\bf Jacobi field
takes the place of the oscillators of gauge fields 
and the sectional curvature plays the role
of ``mass'' of these oscillators}. 

The general properties of the Jacobi equation gives
us a possibility to study chaotic behavior in quantum
area in the spirit of work \cite{CLP}.
\vskip 1cm
ACKNOWLEDGEMENTS
\vskip .2cm
I sincerely thank Larry Horwitz for 
useful discussions and interesting proposales.


\vskip .5cm

\begin{thebibliography}{99}
\bibitem{Dirac1}
P. Dirac, Proc. Roy. Soc.A, {\bf 114} 243 (1927).
\bibitem{Dirac2}
P.Dirac, {\it Lectures on Quantum Field Theory},
(Published by Belfer Graduate School of Science
Yeshiva Universiry, New York, 1967).
\bibitem{Weinberg}
S.Weinberg,{\it The Quantum Theory of Fields},\\
(Cambridge University Press, 1996).
\bibitem{TFD}
H.Umezawa and H.Matsumoto, M. Tachiki,{\it Thermo Field 
Dynamics and Condensed States}, (North-Holland Publishing
Company, Amsterdam-New York-Oxford, 1982).
\bibitem{Barut}
A.O.Barut, Found.Phys. {\bf 24}, (11) 1571 (1994).
\bibitem{Dirac3}
P.Dirac, Proc. Roy. Soc.A, {\bf 268}  57 (1962).
\bibitem{Dirac4}
P.Dirac, Proc. Roy. Soc.A, {\bf 136}  453 (1932).
\bibitem{Le1}
P.Leifer, The Nonlinear Quantum Gauge Theory--
Superrelativity, \\
Preprint gr-qc/9704054.
\bibitem{Le2}
P.Leifer, Inertia as ``Threshold of Elasticity'' of
Quantum States, \\
Preprint gr-qc/9706056.
\bibitem{Le3}
P.Leifer, Int.J.Theor.Phys, in press.
\bibitem{Le4}
P.Leifer, Dynamical Spacetime and the Curvature of Projective State Space \\
Preprint gr-qc/9711059.
\bibitem{Le5}
P.Leifer, Found.Phys. {\bf 27}, (2) 261 (1997).
\bibitem{Horwitz}
L.P.Horwitz, Time and the Evolution of States in
Relativistic Classical and Quantum Mechanics \\
Preprint hep-ph/9606330.
\bibitem{KN}
S. Kobayashi and K. Nomizu, {\it Foundations of Differntial Geometry,vol.II},\\
(Interscience Publishers, New York-London-Sydney, 1969).
\bibitem{WZ}
F.Wilczek and A.Zee, Phys.Rev.Lett. {\bf 52}, (24) 2111
(1984).
\bibitem{Milnor}
J.Milnor, {\it Morse Theory}, (Princeton, Princeton
University Press, 1963).
\bibitem{CLP}
L.Casetti, R.Livi, M.Pettini, Phys.Rev.Lett. {\bf 74},
(3) 375 (1995).
\end{thebibliography}
\end{document}